\documentclass[%
reprint,
 amsmath,amssymb,
 aps,
]{revtex4-1}
\usepackage{epsfig,epsf,rotating}
\usepackage{dcolumn}% Align table columns on decimal point
\usepackage{bm}% bold math
\usepackage{graphicx}% Include figure files
\begin{document}

\preprint{APS/123-QED}

\title{Systematic studies of hadron production spectra in collider
experiments}% Force line breaks with \\

\author{\firstname{A.~A.}~\surname{Bylinkin}}
 \email{xander-snz@rambler.ru}
\affiliation{%
 Institute for Theoretical and Experimental
Physics, ITEP, \\
Moscow, Russia
}%
\author{\firstname{A.~A.}~\surname{Rostovtsev}}
 \email{rostov@itep.ru}
\affiliation{%
 Institute for Theoretical and Experimental
Physics, ITEP, \\
Moscow, Russia
}%

\date{\today}% It is always \today, today,

\begin{abstract}
A shape of invariant differential cross section for hadron
production as function of transverse momentum is analysed.
The systematic analysis of the available data demonstrates a need for
a modification of
the parameterization traditionally used to approximate the
measured spectra. The properties of the new proposed parameterization
are discussed.
\end{abstract}

\pacs{Valid PACS appear here}% PACS, the Physics and Astronomy
\maketitle

%\tableofcontents

\section{Introduction}
    There exists a large body of experimental data on hadron production
in high energy
proton-proton,
photon-proton, photon-photon and heavy ion
collisions~\cite{ISR,SppS0209,UA2,UA1,CDF,CDF2009,H1gp,ZEUSgp,L3,H199,OPAL,H1DIS,ZEUSDIS,RHIC,CMS}.
The spectra of hadrons produced in these
collisions are characterized by an exponential behavior as function of
transverse energy
$(E_T)$ for the
bulk of produced hadrons, which populate the low $E_T$ part of the
spectra. This
behavior resembles the Boltzmann-like
spectrum in classical thermodynamics. The exponential shape of the
spectra changes
to a
power law for high $E_T$ hadrons. This change is traditionally
interpreted as an onset
of the
perturbative QCD regime of  hadron production. These features of the
spectra shape are found
to be
universal for any type of colliding particles. Therefore, it is tempting
to find one
universal smooth functional form, which describes the spectra of
produced hadrons
in the whole available $E_T$
range for different energies and types of colliding particles. The
parameters of such
universal functional form
 are expected to vary for different collision
energies, types of colliding particles and types of produced hadrons.
A study of the variations of these parameters
provides a unique information on the hadron production dynamics.

In the present paper the experimentally measured inclusive spectra of
long-lived charged particles (mainly charged
pions) produced at central rapidities in the colliding particles
center of mass system are analyzed.
Further on it is assumed all charged particles being pions for
simplicity. The present analysis is based
on the published hadron production measurements made with
$pp-$collisions
at ISR~\cite{ISR} and LHC~\cite{CMS},
$p\overline{p}-$collisions at $Sp\overline{p}S$~\cite{SppS0209, UA2,
UA1}
and Tevatron~\cite{CDF, CDF2009}, $AuAu-$collisions with
different centralities at RHIC~\cite{RHIC},
real $\gamma\gamma-$collisions at LEP~\cite{L3, OPAL}
 and $\gamma{p}-$interactions with different values of
photon virtuality~$(Q^2)$ at HERA.
The HERA data additionally allow to consider two different regimes of
$\gamma{p}-$interactions:
photoproduction at low values of $Q^2$~\cite{H1gp, ZEUSgp, H199}
and Deep Inelastic Scattering (DIS) for high $Q^2$ values~\cite{H1DIS,
ZEUSDIS}.
The data for
all these inclusive differential cross section measurements have been
taken with a minimum
bias trigger conditions and at
center of mass energy $(\sqrt{s})$ ranging from 23 to 2360 GeV.

\section{Spectrum Analysis}

\begin{figure}[h]
\includegraphics[width = 8cm]{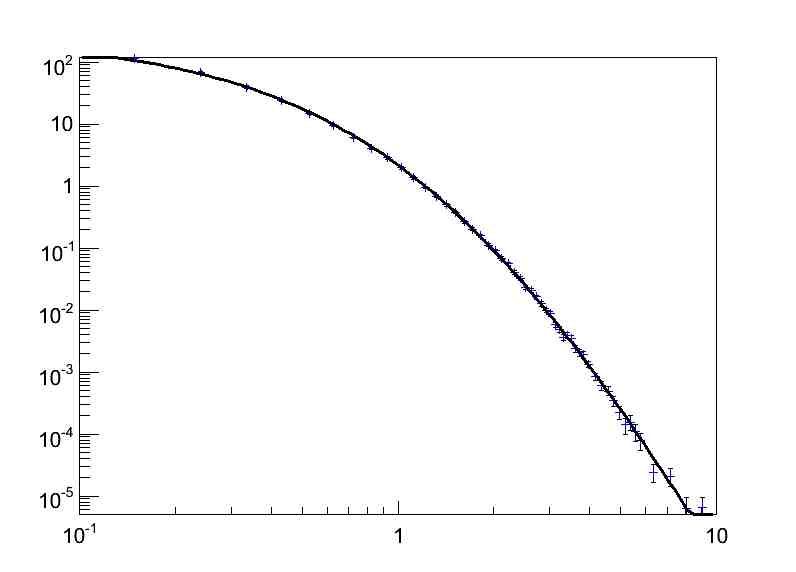}
%\epsfig{file=001.eps}
% bburx=500,
% bbury=543,bbllx=69,bblly=0,height=8.0cm,width=7.0cm,angle=0}
%\put(-148,63){\large\bf $h^+ + h^-$}
%\put(-40,185){\Large\bf a}
%\put(-150,110){\Large UA1}
%\put(-165,85){\small\bf $\sqrt{s}=560~GeV$}
\put(-70,-5){\large\bf $E_T~[GeV]$}
\put(-245,40){\begin{sideways}\large\bf
$E\left.\frac{d^3\sigma}{d^3P}\right|_{y=0} [\mu{b}/GeV^2]$\end{sideways}}
\caption{\label{fig:01} A typical charged particle spectrum fitted using the Tsallis-type function~(\ref{eq:Tsallis}).}
\end{figure}

    A typical charged particle spectrum as function  of  transverse
energy is shown in
Fig~\ref{fig:01}.

This spectrum is fitted using the Tsallis-type function~\cite{Tsallis}
\begin{equation}
\label{eq:Tsallis}
\frac{d\sigma}{P_T d P_T} = \frac{A}{(1+\frac{E_T}{T\cdot n})^n},
\end{equation}
where  $P_T$ is transverse momentum of the produced particle, $E_T =
\sqrt{P_T^2 + M^2}$
with M equal to the pion mass. The parameterization~(\ref{eq:Tsallis})
has only three free
parameters: $A, T$ and $n$. While $A$ is an overall normalization, the
$T$ and $n$ carry
important
information on the hadron production dynamics. Since for low $E_T$
values the
parameterization
~(\ref{eq:Tsallis}) is reduced to the Boltzmann exponent  $\sim~exp
{(-E_T/T)}$, the
parameter $T$ is a QCD
analogy to a temperature in classical thermodynamics. Note, this analogy
isn't straight
forward, however.
\begin{figure*}[!]
\includegraphics[width = 8cm]{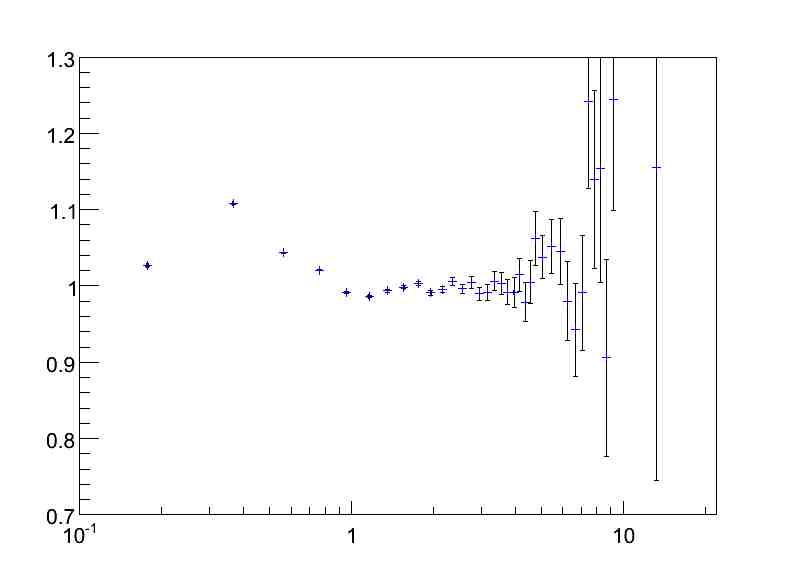}
\put(-80,-5){\large\bf $E_T~[GeV]$}
\put(-230,40){\begin{sideways}\large\bf
$Data/Fit~Ratio$\end{sideways}}
\put(-200,130){\large\bf\ $a)$}
\quad
\includegraphics[width = 8cm]{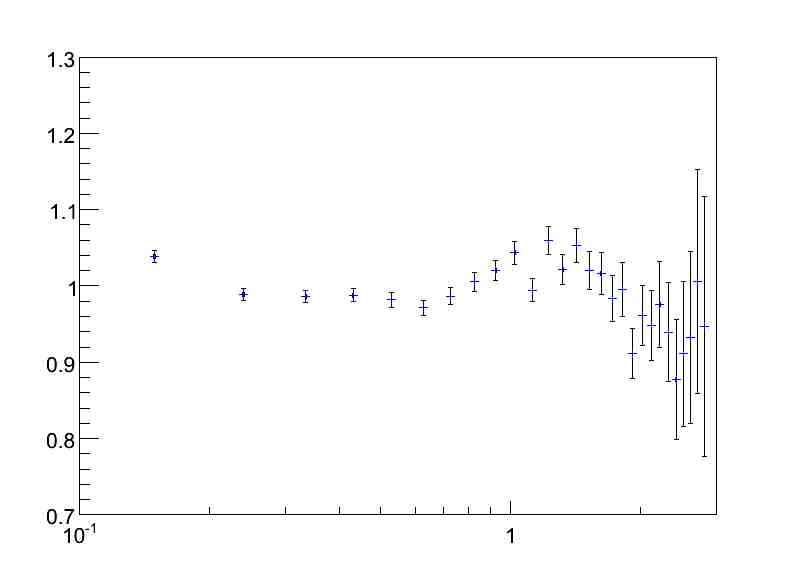}
\put(-80,-5){\large\bf $E_T~[GeV]$}
\put(-230,40){\begin{sideways}\large\bf
$Data/Fit~Ratio$\end{sideways}}
\put(-200,130){\large\bf\ $b)$}
%\caption{\label{fig:02} Systematic defects of the fit using the Tsallis-type function~(\ref{eq:Tsallis}).
%Fig(a): $p\overline{p}$ collisions at $\sqrt{s}=640~GeV$~\cite{UA1}.\\
%Fig(b): RHIC $Au-Au$ collisions at $\sqrt{s}=200~GeV/N$~\cite{UA2}.}

\quad
\includegraphics[width = 8cm]{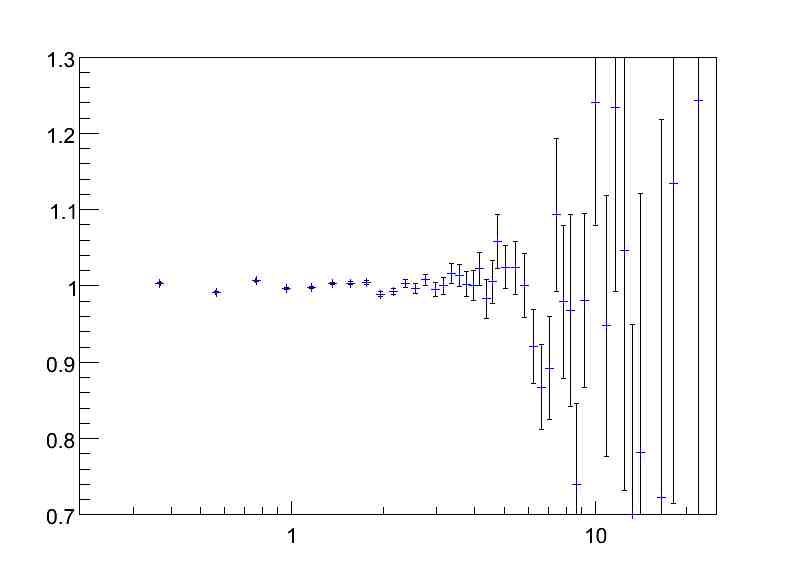}
\put(-80,-5){\large\bf $E_T~[GeV]$}
\put(-230,40){\begin{sideways}\large\bf
$Data/Fit~Ratio$\end{sideways}}
\put(-200,130){\large\bf\ $c)$}
\quad
\includegraphics[width = 8cm]{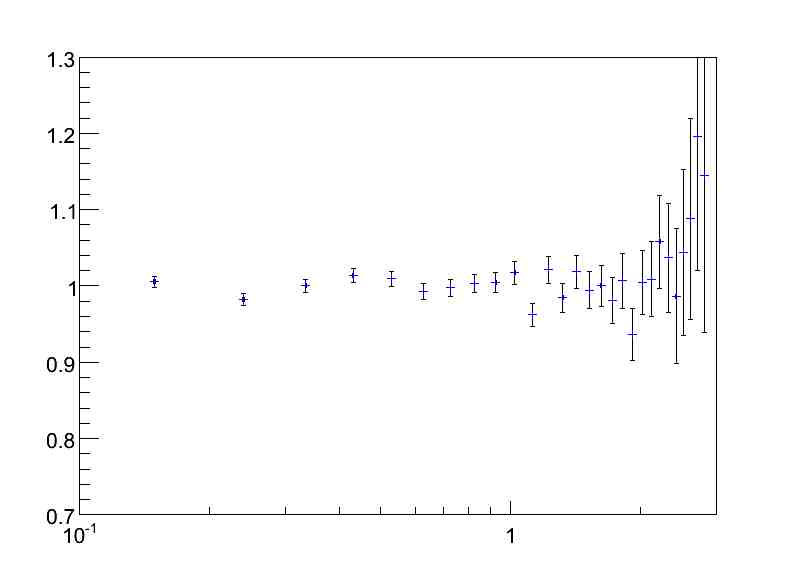}
\put(-80,-5){\large\bf $E_T~[GeV]$}
\put(-230,40){\begin{sideways}\large\bf
$Data/Fit~Ratio$\end{sideways}}
\put(-200,130){\large\bf\ $d)$}
\caption{\label{fig:02} A ratio of the data to the Tsallis-type~(\ref{eq:Tsallis})(a,b) and the modified ~(\ref{eq:exppl})(c,d) fit functions in $p\overline{p}$ collisions at $\sqrt{s}=630~GeV$~\cite{UA1} (a,c) and RHIC $Au-Au$ collisions at $\sqrt{s}=200~GeV/N$~\cite{RHIC} (b,d).}
%Fig(a, c): $p\overline{p}$ collisions at $\sqrt{s}=640~GeV$~\cite{UA1}.\\
%Fig(b, d): RHIC $Au-Au$ collisions at $\sqrt{s}=200~GeV/N$~\cite{UA2}.}
\end{figure*}
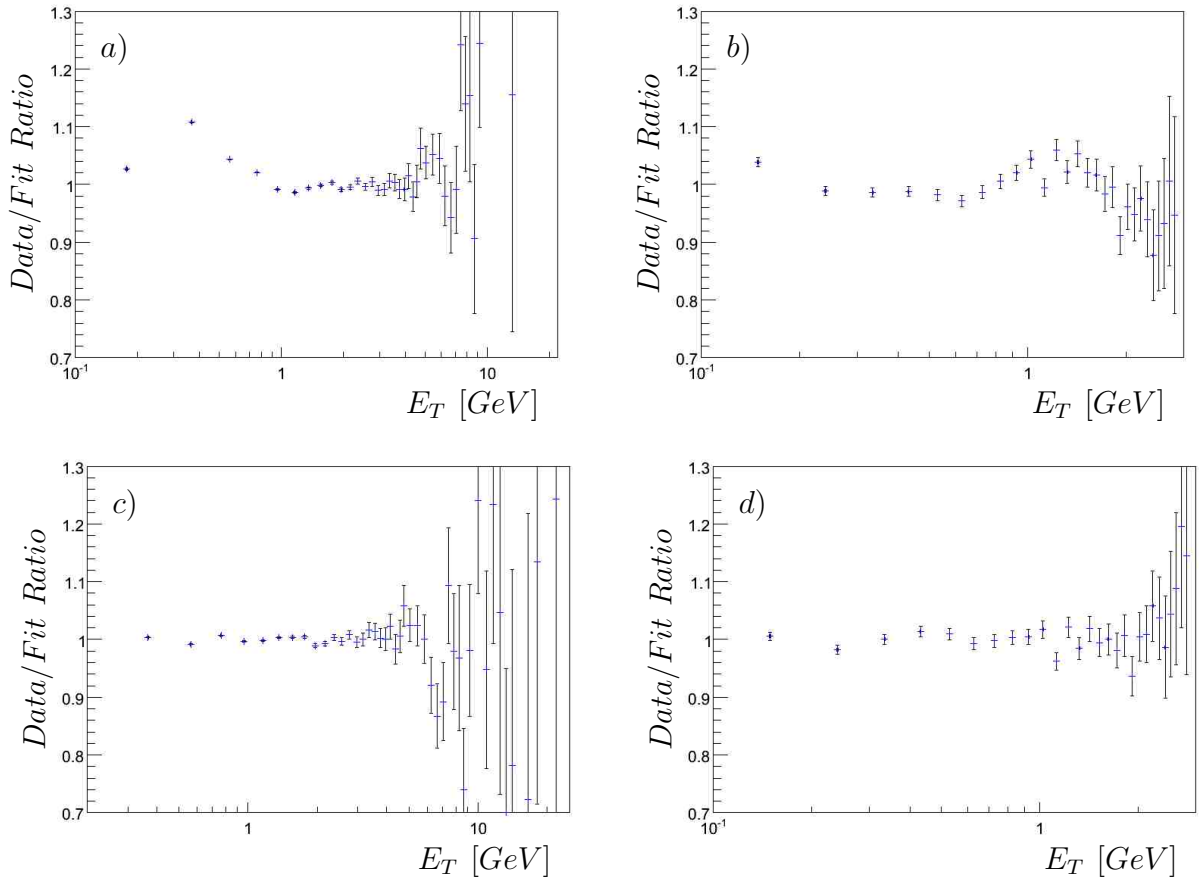
  In perturbative QCD calculations the value of the parameter $n$
depends on
the shape of structure functions and partonic content of the colliding
particles as well as on the differential cross sections of parton-parton
interactions. As it is demonstrated
in Fig~\ref{fig:01}, the economic Tsallis-type parameterization~(\ref{eq:Tsallis})
provides a good
overall
description of the spectrum, thus making the Tsallis-type function
broadly used to fit the recent measurements at RHIC and LHC.%

\section{Modification of the Tsallis function}

\subsection{Systematic defects of the fit}

    However, a closer look at the fit shown in Fig~1 discloses
systematic defects of the fit. In order
to see these defects it is convenient to plot a ratio of the data to the
fit function. Two
such
ratios are shown in Fig~\ref{fig:02}(a,b) for the data sets provided by the UA1
experiment with
$p\overline{p}$ collisions at
$\sqrt{s}=630~GeV$~\cite{UA1}~(Fig~\ref{fig:02}a)
and by RHIC $Au-Au$ collisions at $\sqrt{s}=200~GeV/N$~\cite{RHIC}~(Fig~~\ref{fig:02}b). On
both plots one observes dips and bumps around or above
$1~GeV$ scale.

 The observed difference
between the shapes of data spectra and function~(\ref{eq:Tsallis}) is
also
typical for other available data sets, not shown here. This observation
indicates that the true shape of the measured spectra doesn't follow
exactly the  Tsallis-type parameterization~(\ref{eq:Tsallis}). Moreover,
a larger mismatch
between the
shapes of the data spectrum and fit function~(\ref{eq:Tsallis}) is found
for $J/\Psi$
production at the Tevatron~\cite{CDFJPSI}.
In the following an attempt is made to find a function which
approximates the
shape of the inclusive hadron production spectra better than the Tsallis
parameterization.

\subsection{Search for the new fit function}

   To construct a new fit function a combination of two functional forms
has been used: the
exponential
and power-law.  The both functional forms must be the functions of
scalars
$P_T^2$ or
$E_T^{kin} = E_T -
M$. This choice of these variables is very convenient since the both
$P_T^2$ and $E_T^{kin}$ vary from zero to the
corresponding
kinematical limits. Non-true scalars like $P_T$ or $P_T^3$ have not been
considered here. Finally, the
parameterization
\begin{equation}
\label{eq:exppl}
\frac{d\sigma}{P_T d P_T} = A_e\exp {(-E_T^{kin}/T_e)} +
\frac{A}{(1+\frac{P_T^2}{T^{2}\cdot n})^n},
\end{equation}
turned out to be in a good agreement with the data. The new ratios
between data and fit
function~(\ref{eq:exppl})
are shown in Fig~\ref{fig:02}(c,d) for the considered above UA1 and RHIC data sets
correspondingly. Fig~\ref{fig:02}(c,d) demonstrates a significant improvement of the
quality of the spectra shape approximation by function~(\ref{eq:exppl})
with respect to that using the Tsallis-type function~(\ref{eq:Tsallis}).
In the following
we present the arguments why this improvement is not a trivial result
of increasing the number of free parameters in the fit function.

\section{Correlation between the parameters}

\begin{figure}[h]
\includegraphics[width = 8cm]{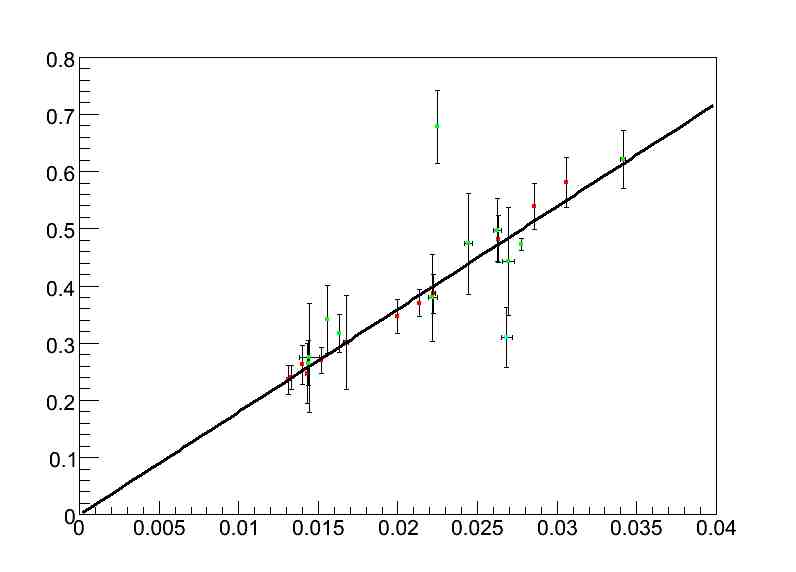}
\put(-80,-5){\large\bf $T_{e}^{2}~[GeV^{2}]$}
\put(-230,90){\begin{sideways}\large\bf
$T^{2}~[GeV^{2}]$\end{sideways}}
\caption{\label{fig:04} A correlation between
the parameters $T_e^{2}$ and $T^{2}$}
\end{figure}

    The most surprising feature of the new parameterization~(\ref{eq:exppl})
is a strong
correlation between
the parameters $T_e$ and $T$. The dependence of fitted values of the
parameter $T^{2}$ versus
$T_e^{2}$ for charged
particles produced in $pp$, $p\overline{p}$ and $Au-Au$ collisions is
shown in Fig~\ref{fig:04}.

This dependence is approximated
by a linear function. This provides an
additional constraint for the parameterization~(\ref{eq:exppl}) and
therefore reduces a number
of free parameters used in the fit to the data.
Though the physical origin of the observed correlation is not quite
clear,
the new constraint helps to
minimize uncertainties of the parameter values obtained from the fits.
This constraint
 will be used further on through this paper.

\section{A toy model interpretation}

   The form of the parameterization~(\ref{eq:exppl}) has a simple
toy-model interpretation. Within this toy-model a small fraction of
hadrons are produced
directly in
parton-parton interactions of the colliding particles. As required by the
perturbative QCD
the spectrum of particles produced in parton-parton interactions is
described
by a power law distribution. The rest bulk of irradiated hadrons
represents a
quasi-thermolized hadronic gas produced with a characteristic
temperature $T_e$.
The spectrum of hadrons in this gas has the Boltzman
exponential shape.

\subsection{Contributions of the exponential\\ and power law
terms}
\begin{figure}[h]
\includegraphics[width = 8cm]{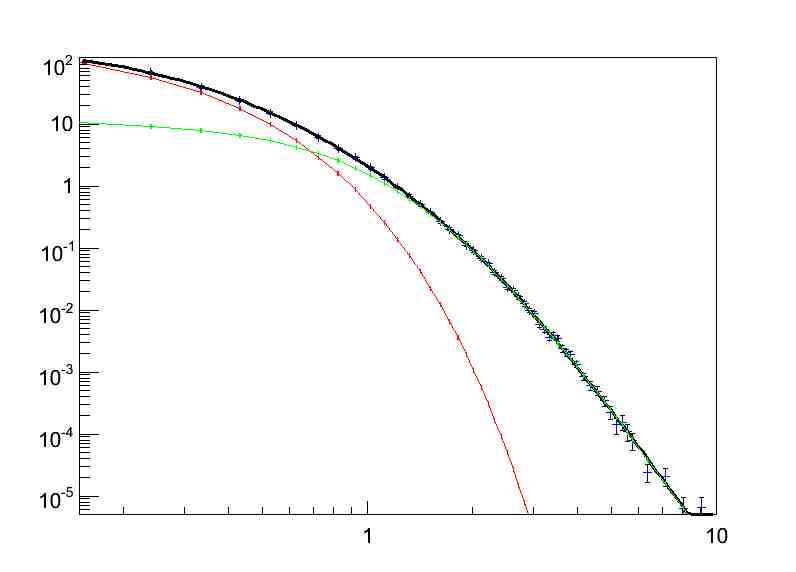}
\put(-80,-5){\large\bf $E_T^{kin}~[GeV]$}
\put(-245,40){\begin{sideways}\large\bf
$E\left.\frac{d^3\sigma}{d^3P}\right|_{y=0} [\mu{b}/GeV^2]$\end{sideways}}
\caption{\label{fig:05} The contributions of the exponential (red) and the power law (green) terms to the spectrum of charged particles produced in $p\overline{p}$ collisions.}
\end{figure}

The contributions of the exponential and power law
terms
of the parameterization~(\ref{eq:exppl}) to
the typical spectrum of charge particle produced in $pp$
collisions are shown separately on Fig~\ref{fig:05}.
The relative contribution of these terms is characterized by a ratio $R$ of the exponential to power law terms integrated over $P_t^{2}$:
\begin{equation}
R = \frac{A_e(2m + 2T_e)(n-1)}{18.11A{\cdot}n{\cdot}T_e}
\end{equation}
For the spectrum in Fig~\ref{fig:05}
the exponential term dominates, while the power law term contributes
at the level of about $20\%$.

\begin{figure*}[!]
\includegraphics[width = 8cm]{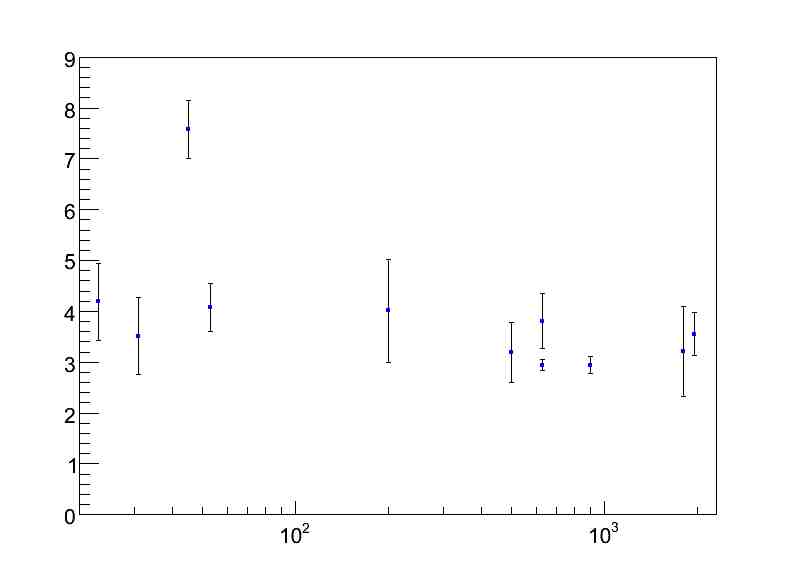}
\put(-80,-5){\large\bf $\sqrt{s}~[GeV]$}
\put(-230,140){\large\bf $R$}
\put(-80, 130){\large\bf $a)$}
\quad
\includegraphics[width = 8cm]{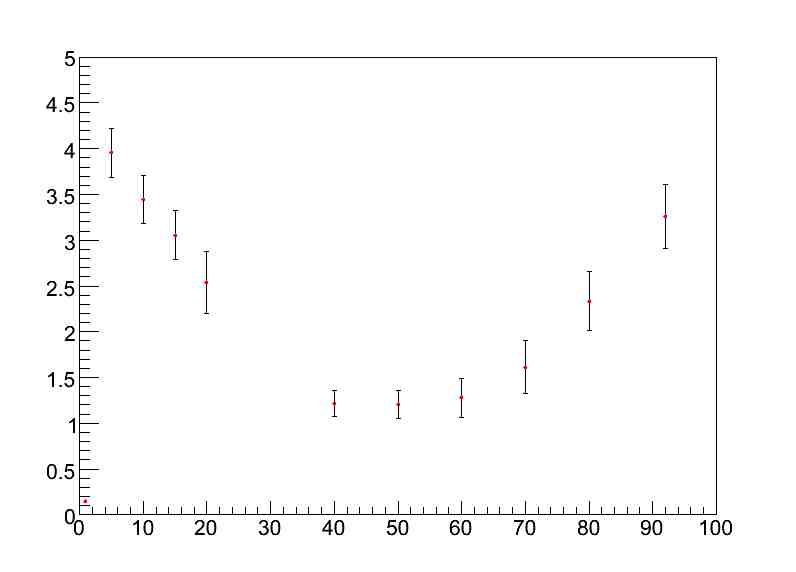}
\put(-80,-5){\large\bf $Centrality [\%]$}
\put(-230,140){\large\bf $R$}
\put(-80, 130){\large\bf $b)$}
\caption{\label{fig:06} The ratio $R$ of the exponential to power law contributions to the
parameterization~(\ref{eq:exppl})for $p\overline{p}$ and $pp$ collisions as function of $\sqrt{s}$ (a)
and for $Au-Au$ interactions at RHIC as function of centrality (b).
}
\end{figure*}
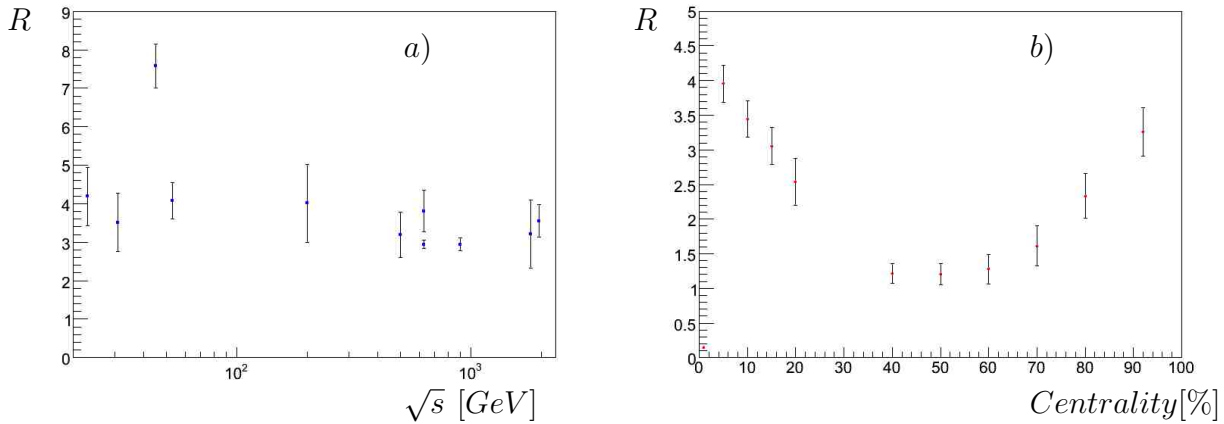

The ratio $R$ for the inclusive charged particle spectra for
 $p\overline{p}$ and $pp$ collisions as function of $\sqrt{s}$ is shown
in Fig~\ref{fig:06}a.
  Interestingly, this ratio is almost independent of the
collision energy and equals to about $4$.
In Fig~\ref{fig:06}b the ratio $R$ is shown for $Au-Au$
interactions at RHIC
as function of
centrality of the heavy ions collision.
As seen on Fig~\ref{fig:06}b the relative
contribution of the
exponent term reaches minimum values at medium centralities of heavy
ion collisions.

\begin{figure}[h]
\includegraphics[width = 8cm]{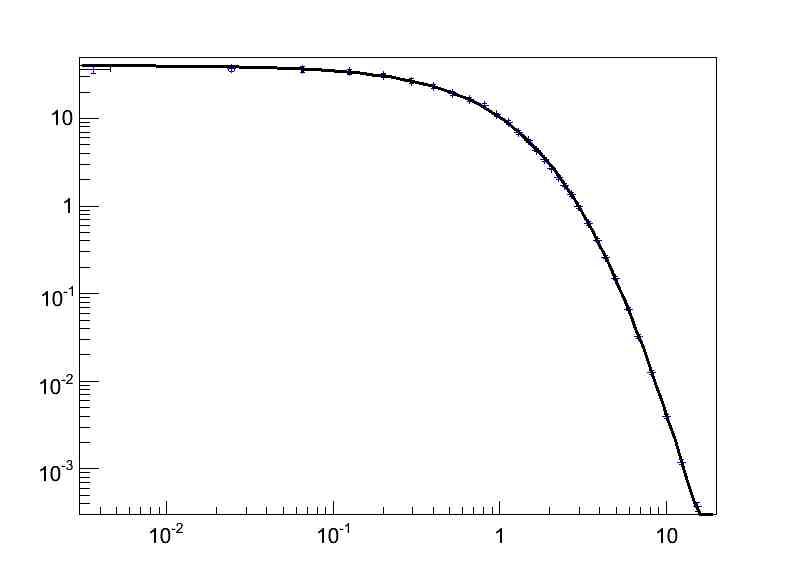}
\put(-80,-5){\large\bf $E_T^{kin}~[GeV]$}
\put(-245,40){\begin{sideways}\large\bf
$E\left.\frac{d^3\sigma}{d^3P}\right|_{y=0} [\mu{b}/GeV^2]$\end{sideways}}
\caption{\label{fig:07} The spectrum of $J/\Psi$ produced in
high energy $p\overline{p}$ collisions at Tevatron~\cite{CDFJPSI}.}
\end{figure}

\begin{figure}[h]
\includegraphics[width = 8cm]{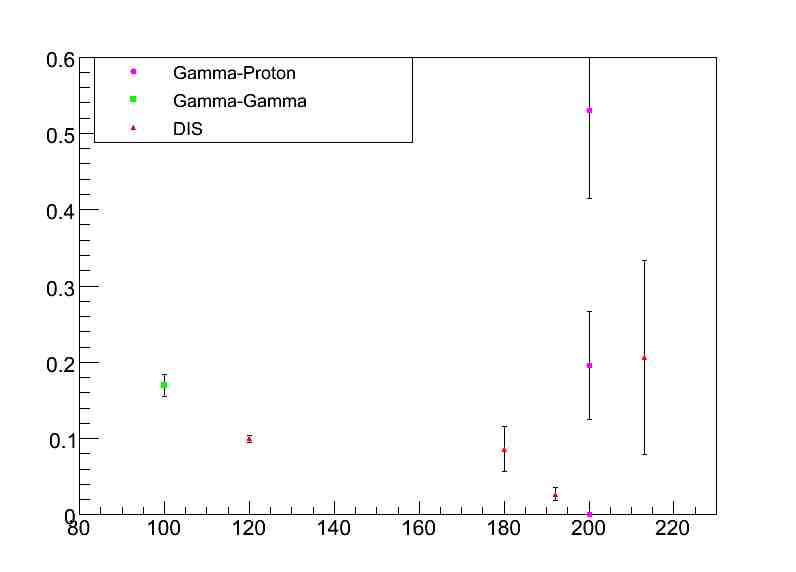}
\put(-80,-5){\large\bf $\sqrt{s}~[GeV]$}
\put(-230,140){\large\bf $R$}
\caption{\label{fig:08} The ratio $R$ of the exponential to power law contributions to the
parameterization~(\ref{eq:exppl}) in the high energy
DIS, photoproduction and $\gamma\gamma$ collisions.}
\end{figure}

On the other hand, the QCD partonic interaction must describe any hard
scattering process
like the heavy quark production, or high $Q^2$ DIS. Indeed,
 as shown in Fig~\ref{fig:07} the spectrum of heavy $J/\Psi$ quarkonium produced in
high
energy $p\overline{p}$ collisions at Tevatron~\cite{CDFJPSI}
follow the pure power law distribution and has no room for an
exponential
term of
the parameterization~(\ref{eq:exppl}).
 In addition, in the high energy
DIS, photoroduction and
$\gamma\gamma$ collisions the power law term of the new proposed
parameterization~(\ref{eq:exppl}) dominates the produced particle
spectra as shown in Fig~\ref{fig:08}.
Thus, only the inclusive spectra of
 charged particles produced in pure baryonic collisions require a
substantial contribution
of the Boltzman-like exponential term. We found it to be particularly
interesting that the
interactions of real photons and protons with real photons (high energy
photoproduction)
have practically no Boltzman-like thermolized hadronic final states.

\section{Map of Parameters}

\begin{figure*}
\includegraphics[width = 18cm]{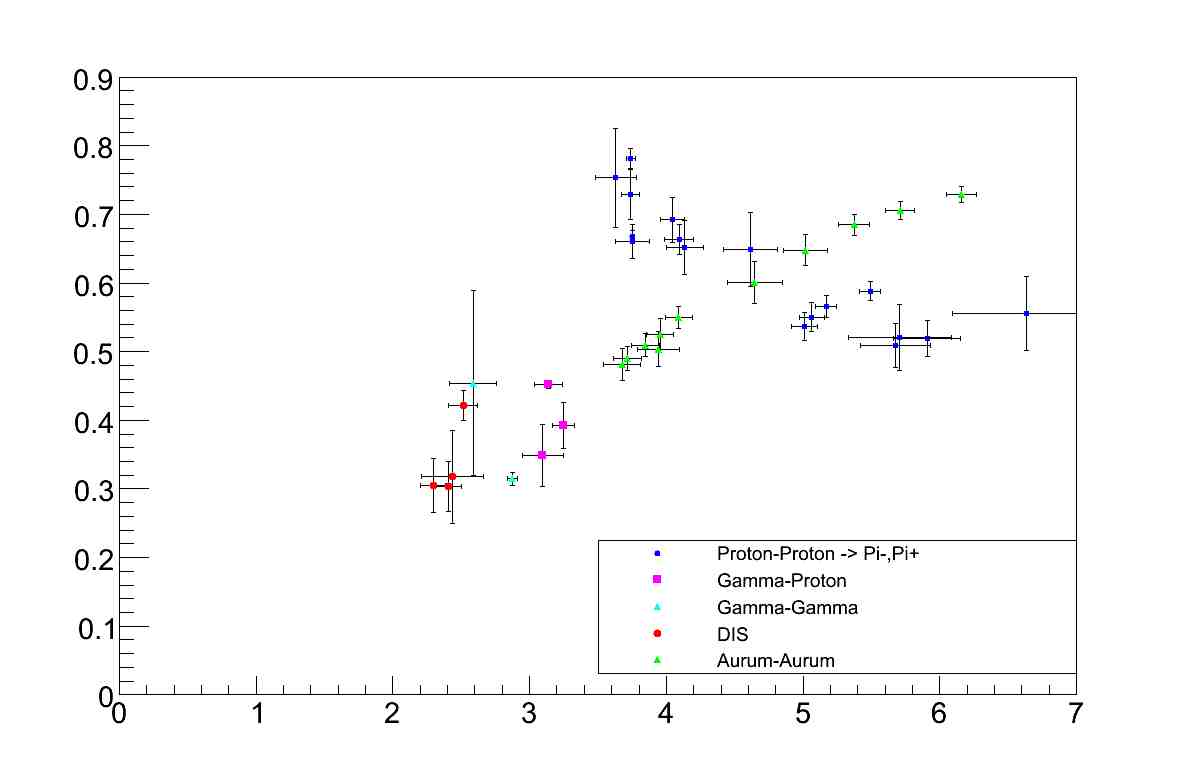}
\put(-80,0){\Large\bf $n$}
\put(-500,240){\begin{sideways}\large\bf
$T^{2}~[GeV^{2}]$\end{sideways}}
\caption{\label{fig:09} Map of the parameters $T$ and $n$ for proton-(anti)proton, heavy ion,
gamma-proton and gamma-gamma collision at different energies.}
\end{figure*}

 Finally, a map
of the parameters $T$ and $n$ for proton-(anti)proton, heavy ion,
gamma-proton and gamma-gamma
collision at
different energies is displayed in Fig~\ref{fig:09}.
 There are two clearly
distinct trends
seen in Fig~\ref{fig:09}. The $pp$ and $p\overline{p}$ collision data show an
increase
of the $T-$parameter and decrease of the
$n-$parameter with collision energy $\sqrt{s}$ increasing.
The second trend, where the values of both parameters the $T$ and $n$
increase, is defined
mainly by the RHIC $Au-Au$
collision data at $\sqrt{s}=200~GeV$ per nucleon. In this case a
simultaneous increase of
the $T$ and $n$ values corresponds to an increase of the
centrality (or charged multiplicity) of heavy ion collisions.
 Surprisingly, the  both trends cross each other at medium
centralities corresponding to the minimum bias $Au-Au$ collisions and
$p\overline{p}$ interactions with energy of $\sqrt{s}=200~GeV$.
Naively one could expect the single $p\overline{p}$ interaction has more
similarity to the very peripheral single nucleon-nucleon interactions.
Contrary to that naive expectation, the Deep
Inelastic Scattering (DIS), $\gamma{p}$ and $\gamma\gamma$ interaction
data
with  $\sqrt{s}$ ranging
approximately from $100~GeV$ to $200~GeV$
belong to the second trend shown in Fig~\ref{fig:09} and are
located on the
parameter map (Fig~\ref{fig:09}) nearby very peripheral heavy ion interactions at
about the same
collision energy per nucleon.

\section{Shape of the spectrum at high $P_t$}
\begin{figure}[h]
\includegraphics[width = 8cm]{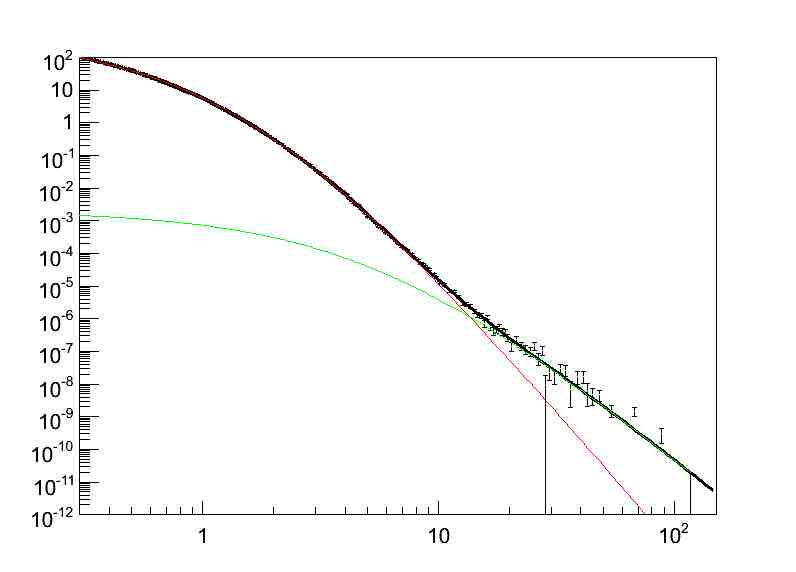}
\put(-80,-5){\large\bf $E_T^{kin}~[GeV]$}
\put(-245,40){\begin{sideways}\large\bf
$E\left.\frac{d^3\sigma}{d^3P}\right|_{y=0} [\mu{b}/GeV^2]$\end{sideways}}
\caption{\label{fig:10} CDF data~\cite{CDF2009} fitted by two power law functions with different free parameters.}
\end{figure}

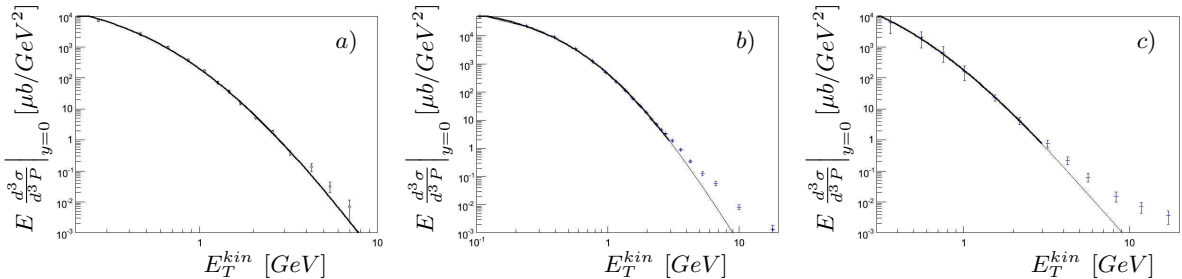
\begin{figure*}[!]
\includegraphics[width = 5cm]{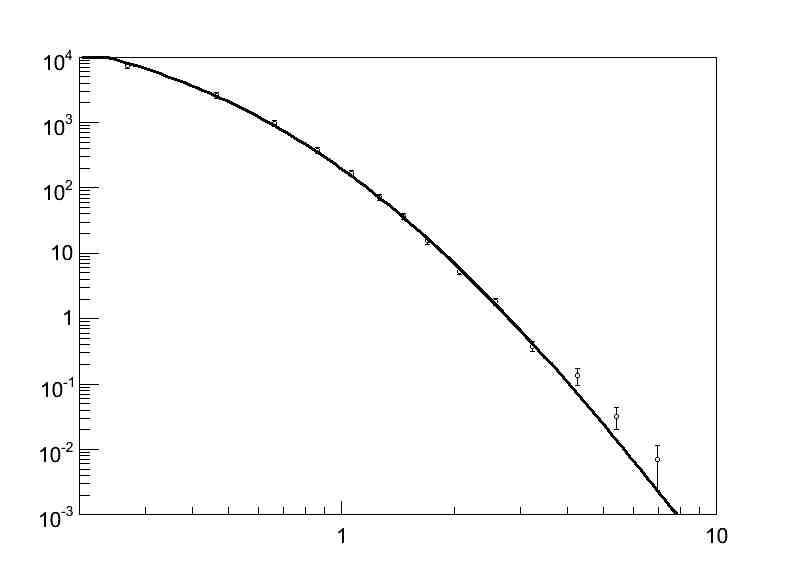}
\put(-80,-5){\bf $E_T^{kin}~[GeV]$}
\put(-155,10){\begin{sideways}\bf
$E\left.\frac{d^3\sigma}{d^3P}\right|_{y=0} [\mu{b}/GeV^2]$\end{sideways}}
\put(-30, 80){\bf $a)$}
\quad
\includegraphics[width = 5cm]{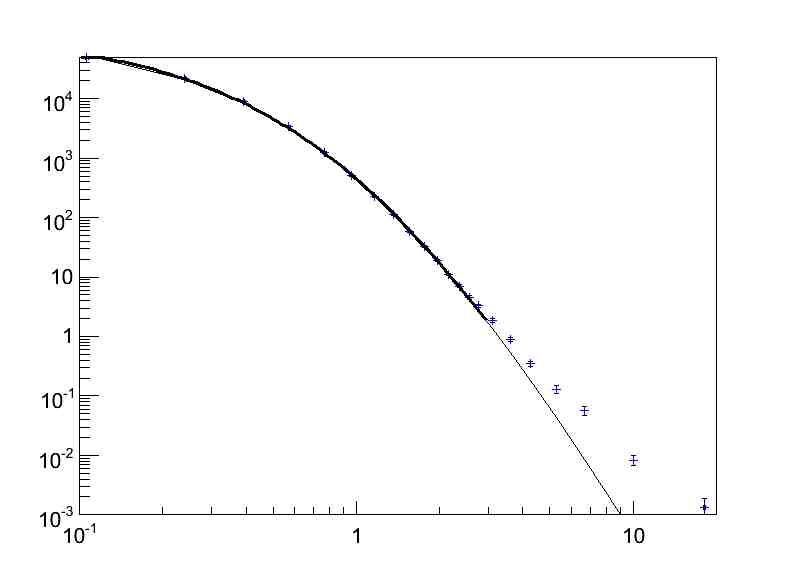}
\put(-80,-5){\bf $E_T^{kin}~[GeV]$}
\put(-155,10){\begin{sideways}\bf
$E\left.\frac{d^3\sigma}{d^3P}\right|_{y=0} [\mu{b}/GeV^2]$\end{sideways}}
\put(-30, 80){\bf $b)$}
\quad
\includegraphics[width = 5cm]{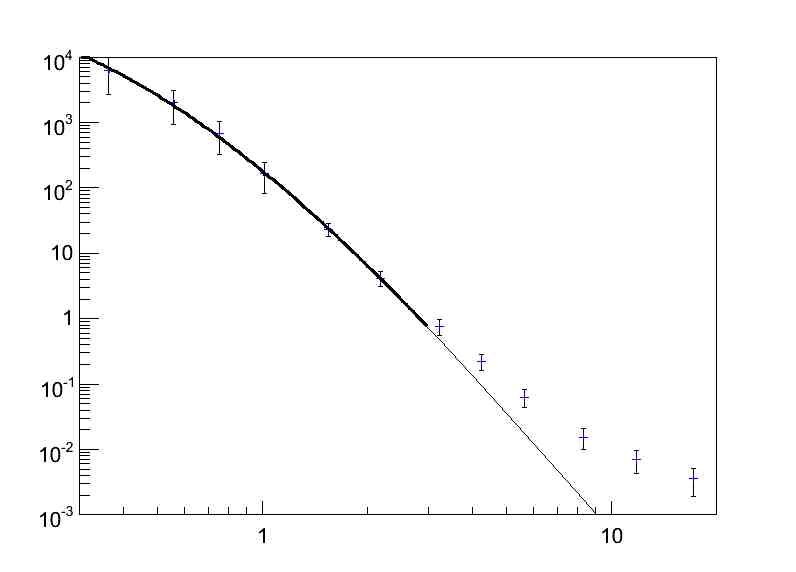}
\put(-70,-5){\bf $E_T^{kin}~[GeV]$}
\put(-155,10){\begin{sideways}\bf
$E\left.\frac{d^3\sigma}{d^3P}\right|_{y=0} [\mu{b}/GeV^2]$\end{sideways}}
\put(-30, 80){\bf $c)$}
\caption{\label{fig:11} Excess over the simple power law shape for $\gamma{p}$~\cite{H1gp}(a) and $\gamma\gamma$~\cite{L3, OPAL}(b,c) interactions.
}
\end{figure*}

The proposed new parameterization~(\ref{eq:exppl}) describes the data
well at low and
medium values of the transverse momentum of the produced particles.
However, as it
was recently shown
by the CDF measurements, in $p\overline{p}$ interaction the inclusive
particle
spectrum has a significant excess over the simple power law shape for
particles
produced with
$P_T > 10~GeV/c$~\cite{CDF2009}. In Fig~\ref{fig:10} it is demonstrated, that
observed excess
is well described by adding a second power law function with somewhat
lower value of
the exponent $n$.
This phenomenon, which has no consistent explanation by now, seems to be
found earlier in $\gamma{p}$ and $\gamma\gamma$ interactions (shown in
Fig~\ref{fig:11}).

Though, in these types of collisions with at least one photon involved
the excess of the data spectra over a single power law function is
visible
already for particles with $P_T > 3 GeV/c$.

%\section{Dips in the spectra}

%Additional interesting feature of the inclusive particle production is a
%small
%dip on a smooth shape of the spectrum at
%high $P_T$ $p\overline{p}$ high energy interactions. This dip is clearly
%visible
%in Fig~(\ref{fig:12}), where the ratios of the data to the simple power law
%fit function for two data sets
%are shown.
%\begin{figure*}[!]
%\includegraphics[width = 8cm]{005}
%\put(-80,-5){\large\bf $E_Tkin~[GeV]$}
%\put(-230,40){\begin{sideways}\large\bf
%$Data to Fit Ratio$\end{sideways}}
%\quad
%\includegraphics[width = 8cm]{021}
%\put(-80,-5){\large\bf $E_Tkin~[GeV]$}
%\put(-230,40){\begin{sideways}\large\bf
%$Data to Fit Ratio$\end{sideways}}
%\caption{\label{fig:12} A small dip on a smooth shape of the spectrum at
%high $P_T$ $p\overline{p}$ high energy interactions).
%}
%\end{figure*}
%These two data sets have been taken at different collision
%energies by
%the UA1~\cite{UA1} and CDF~\cite{CDF2009} experiments. These particular
%measurements
%have been selected for comparison since among all available data only
%these two data sets are extended to the large values of
%produced particle transverse momentum and at the same time have very
%high precision.
%The dips found in the spectra remind the diffractive dips observed in
%elastic particle
%scattering. This new phenomenon, if supported by further measurements,
%poses fundamental questions to the underlying dynamics of hadron
%production at high energies.
%The new LHC high precision data would help to answer these questions in
%the future.

\section{conclusion}

In conclusion we have proposed a new parameterization of the spectrum
shape
of inclusive charged particles produced in high energy collisions.
This new parameterization describes the available experimental data
significantly
 better than the broadly used Tsallis-type parameterization. The
proposed
parameterization is a sum of an exponential (Boltzman-like) and a power
law
(Tsallis-like) terms. The parameters of these two terms turned out to be
strongly correlated.
We observe, that the shapes of the power law terms in minimum bias heavy
ion collisions
and in proton-(anti)proton interactions at the same collision energy are
practically
the same. Additionally, the shapes of the power law terms in the spectra
measured with very
peripheral heavy ion collisions and interactions of high energy
particles with photons
show very close similarity. The difference in size of the exponential
Boltzman-like
contribution to these spectra is mainly responsible for the difference
in the spectra shapes observed in these experiments.

\begin{acknowledgments}
The authors thank professor M.Ryskin for helpful discussions.
This work was partially supported by
Russian Foundation for Basic Research and
the Grant of Helmholtz Association HRJRG-02.
\end{acknowledgments}
%\newpage

%\nocite{*}

\end{document}